\documentclass[superscriptaddress,prc,twocolumn,nofootinbib]{revtex4-2} 
\usepackage{color}

\usepackage{amsmath}
\usepackage{amssymb}
\usepackage{graphicx}
\usepackage{bm}
\usepackage{hyperref}
\usepackage{url}
\hypersetup{colorlinks = true, allcolors = blue}

\usepackage[normalem]{ulem}  

\renewcommand{\vec}[1]{\mbox{\boldmath $#1$}}

\begin{document}

\title{Analysis of a Skyrme energy density functional with deep learning}

\author{N. Hizawa}
\author{K. Hagino}
\affiliation{
Department of Physics, Kyoto University, Kyoto 606-8502,  Japan}
\author{K. Yoshida}
\affiliation{Research Center for Nuclear Physics, Osaka University, 
Ibaraki, Osaka 567-0047 Japan}
\affiliation{RIKEN Nishina Center for Accelerator-Based Science, Wako, Saitama 351-0198, Japan}
\begin{abstract}
Over the past decade, machine learning has been successfully applied in various fields of science.
In this study, we employ a deep learning method to analyze a Skyrme energy density functional (Skyrme-EDF), that is 
a Kohn-Sham type functional commonly used in nuclear physics.
Our goal is to construct an orbital-free functional that reproduces the results of the Skyrme-EDF.
To this end, we first compute energies and densities 
of a nucleus with the Skyrme Kohn-Sham + Bardeen-Cooper-Schrieffer method by introducing a set of external fields. 
Those are then used as training data for deep learning to construct a functional which depends 
only on the density distribution. 
Applying this scheme to the $^{24}$Mg nucleus with two distinct random external fields, we successfully 
obtain a new functional which reproduces the binding energy of the original Skyrme-EDF
with an accuracy of about 0.04 MeV. 
The rate at which the neural network outputs the energy for a given density is about $10^5$--$10^6$ times faster than the Kohn-Sham scheme, 
demonstrating a promising potential for applications to heavy and superheavy nuclei, including the dynamics of fission.
\end{abstract}

\maketitle

\section{Introduction}
Recent progress of deep learning is quite remarkable. 
It has actually gained popularity in various fields 
of science and technology, such as natural language processing, computer vision, 
and speech recognition \cite{BERT, GPT3, stable_diffusion, ViT, Conformer}.
In several fields of physics, such as condensed matter physics, 
a multitude of ideas to utilize machine learning are arising. 
For instance, in Ref. \cite{RS19}, an energy density functional (EDF) for electron systems 
that depends solely on an electron number density was constructed using the method developed in Ref. \cite{MS17}, 
in which an attempt was made with a neural network to predict the solution of a two-dimensional Schr\"odinger equation 
in a random potential.
Other applications have already existed also for a variety of problems, including those in spin systems \cite{SM20} and 
superconducting systems \cite{VC18}.

In contrast, an application of deep learning to nuclear physics has still been in its early stage \cite{SJ21, WP21, WG21, MS22, LM22, SL22, ML22, WR22, VS22, MT23, ZL23, NB23, YZ23, ZY23, KH23, BD22}.
We mention that nuclear physics continues to face numerous unresolved challenges that call for innovative solutions, 
including a description of large amplitude collective motions. Such problems may be solved efficiently by applying the machine learning techniques, 
developed in other fields of physics. 

In particular, the recent application of deep learning 
to the Kohn-Sham type DFT \cite{RS19} mentioned above  
could be readlily applied also to nuclear physics. 
In nuclear physics, phenomenological models for a functional 
have often been employed \cite{BH03}.
The resultant Kohn-Sham type energy density functional (KS-EDF) is not an explicit 
functional of the particle number density only, but is parameterized together with other local densities, 
such as the kinetic energy density, the spin-orbit density, and the pair density when 
considering explicitly the nucleonic superfluidity.
To calculate observables using the KS-EDF, such as the binding energy of a nucleus, 
one needs to solve
a self-consistent differential equation of the same form as that in 
the mean-field theory many times, 
which is computationally expensive especially for heavy systems.
Therefore, it is desirable to develop an orbital-free EDF (OF-EDF) theory that does not depend 
on Kohn-Sham orbitals. Deep learning can be a powerful tool for that purpose \cite{RS19}. 
Such theory will be based on a functional that 
depends sorely on the particle number density. 
Notice that this is totally consistent with  
the original philosophy of the density functional theory (DFT). 

The aim of this paper is to apply the method developed in Ref. \cite{RS19} to a nuclear system and 
construct a deep-learning-based nuclear OF-EDF that reproduces results of the Skyrme-EDF. 
In applying the method of Ref. \cite{RS19} to a nuclear system, one has to take into account 
several aspects that make nuclear systems different from electron systems. 
One obvious difference is that a nucleus is a self-bound attractive system. 
In a electron system without phonons, the only interaction between electrons 
is the repulsive Coulomb force, which causes two electrons to distribute as far as possible. 
In marked contrast, nucleons tend to get closer to each other due to a short-ranged attractive nuclear force, and thus the mechanism which determines the density distribution is quite different between electron and nucleon systems \cite{Naito2021}. 
In addition, for electron systems, the KS-EDFs, which are inspired by the Hartree-Fock method, works well in general.
On the other hand, in nuclear systems, 
superfluidity plays a crucial role in open-shell nuclei, and observables are better explained using a KS-EDF that is inspired by the Hartree-Fock-Bardeen-Cooper-Schrieffer (BCS) or Hartree-Fock-Bogoliubov method rather than by the Hartree-Fock method.
This leads to a technical difference in that a nuclear KS-EDF depends also on the pair density.
It will be intriguing to investigate how well the deep learning method works 
in such attraction-dominated nuclear systems.

The paper is organized as follows.
In Sec. II, we introduce the KS-EDF which we employ, and define a protocol for deep learning.
We also discuss how to generate data sets to train neural networks.
In Sec. III, we carry out the deep learning for the $^{24}$Mg nucleus 
and discuss how well the data sets can be learned.
We then summarize the paper in Sec. IV and discuss future perspectives.

\section{Formulation}
\subsection{Skyrme EDF}
We first introduce a Kohn-Sham type energy density functional (KS-EDF) 
for training on a neural network.
Throughout this study, we consistently employ the following Skyrme-type EDF \cite{Ring_Schuck}:
\begin{equation}
\label{eq:KS-EDF}
    E_{\mathrm{tot}}
     =
    E_{\mathrm{kin}} + E_{\mathrm{int}} + E_{\mathrm{pair}} + E_{\mathrm{CoM}},
\end{equation}
with 
\begin{align}
    E_{\mathrm{kin}}[\tau]
      & =
      \frac{\hbar^2}{2m}\left(1-\frac{1}{A}
      \right)
      \sum_q \int d^3r \,\tau_q(\bm{r}), \\
      E_{\mathrm{int}}[\rho, \tau, \bm{J}]
      & = \int d^3r\biggl\{
      \frac{b_0}{2}\rho^2 - \frac{b'_0}{2}\sum_{q}\rho_q^2 \notag \\ 
      &+ \frac{b_3}{3}\rho^{\alpha+2} - \frac{b'_3}{3}\rho^{\alpha}\sum_{q}\rho_q^2
      +b_1\rho\tau - b'_1\sum_{q}\rho_q\tau_q
      \notag \\
      & \qquad \qquad
      - \frac{b_2}{2}\rho\nabla^2\rho
      + \frac{b'_2}{2}\sum_{q}\rho_q\nabla^2\rho_q 
      \notag \\
      &-b_4\rho\nabla\cdot \bm{J}
      -b'_4\sum_{q}\rho_q\nabla\cdot \bm{J}_q
      \biggr\},
      \\
      E_{\mathrm{pair}}[\rho, \tilde{\rho}]
      & = 
      \sum_{q}\frac{V^{(q)}_0}{4}
      \int d^3 r
      \left\{1-
      \left(\frac{\rho}{\rho_0}\right)^{\gamma}
      \right\}\tilde{\rho}_q^2, 
\end{align}
and
\begin{align}
      E_{\mathrm{CoM}}[\rho]
      &=
      \frac{C}{2}\left(\int d^3r\,z\rho(\bm{r})\right)^2,
      \label{eq:KS-EDF-COM}
\end{align}
where $m$ is the nucleon mass and $A$ is the mass number of a nucleus. 
$E_{\mathrm{kin}},~E_{\mathrm{int}},~E_{\mathrm{pair}}$, and $E_{\mathrm{CoM}}$ are the kinetic energy, the interaction energy, 
the pairing energy, and a cost function for the center-of-mass, respectively. 
$\rho, \tau, \bm{J}$, and $\tilde{\rho}$ are the particle number density, the kinetic density, the spin density, and the pair density, respectively, in which the subscript $q$ refers to neutron or proton. 
Those are defined as
\begin{align}
    \rho(\vec{r})
      & = 2\sum_{q}\sum_{k>0}v_{q, k}^2|\varphi_{q, k}(\bm{r})|^2,
      \\
    \tau(\vec{r})
     & = 2\sum_{q}\sum_{k>0}v_{q, k}^2|\nabla\varphi_{q, k}(\bm{r})|^2,
     \\
    \vec{J(\vec{r})}
    &= 2\sum_{q}\sum_{k>0}v_{q, k}^2
    \varphi^{*}_{q, k}(\bm{r})
    \left(-i\nabla\times\bm{\sigma}\right)
    \varphi_{q, k}(\bm{r}),
    \\
    \tilde{\rho}(\vec{r})
    & = -2\sum_{q}\sum_{k>0}u_{q, k}v_{q, k}|\varphi_{q,k}(\bm{r})|^2,
\end{align}
where $\varphi_{q, k}(\bm{r})$ is the $k$-th Kohn-Sham orbrbital in a spinor form with isospin $q$, and $v_{q, k}^2 = 1-u_{q, k}^2$ is the occupation probability for the $k$-th orbital.
Notice that we take the BCS approximation for the treatment of the pairing correlation. 

In the interaction part of the functional, $b_i$ and $b'_i~(i=1\text{--}4)$ as well as $\alpha$ are the Skyrme parameters.
In this paper, we use the SLy4 parameter set \cite{SLy4} for these parameters.
For simplicity, we ignore the Coulomb interaction, even though 
the entire Coulomb interaction term can be explicitly described as a functional of the proton number density if the 
Slater approximation is introduced to the exchange term. 

For the pairing part, we employ a surface-type functinal of the Density-Dependent Delta-Interaction (DDDI) \cite{DDDI}, 
which contains the parameters 
$V_0^{(q)}, \rho_0$, and $\gamma$. 
In this study, we take $\gamma = 1$ and $\rho_0 = 0.16\,\mathrm{fm}^{-3}$, and determine $V_0^{(q)}$ so that the average pairing gap coincides with the empirical pairing gap, $\Delta_q = 12/\sqrt{A}$ MeV \cite{Bohr_Mottelson1, Ring_Schuck}.
The zero-range pairing interaction has to be supplemented with an energy cut-off. 
In this paper, the sharp cut-off energy of 60 MeV is introduced to the single particle energy of 
the Kohn-Sham orbitals. 
The resultant strengths for the pairing are $V_0^{(n)} = V_0^{(p)}=-683.344$ MeV fm$^3$.

In addition to the ordinal Skyrme EDF, we introduce a functional 
$E_{\mathrm{CoM}}[\rho]$ to fix the center-of-mass position in the $z$ direction.
This is necessary as we introduce external fields (see Sec. II C below) to generate various density distributions. 
By fixing the center-of-mass position, one can prevent a nucleus from localizing around the edges of the box, 
which is useful to generate various deformed states in a small box. 
In this study, we take $0.625\,\mathrm{MeV/fm^2}$ for the value of $C$. 

In this paper, we consider only the $^{24}$Mg nucleus.
This choice of a nucleus is convenient, as this nucleus has equal numbers of protons and neutrons, and thus the proton and the neutron densities coincide to each other when the Coulomb interaction is ignored. 
Furthermore, we impose the axial symmetry and the time-reversal symmetry on the system, 
enabling the local densities to be expressed in the cylindrical coordinates $(r,z)$ \cite{TO03}. 
Notice that Ref. \cite{RS19} also used a two-dimensional EDF for 
electron systems. 
With these simplifications, in principle, the EDF of the system should be able to be expressed solely with the nucleon number density 
$\rho(r, z)$, which can be considered as a monochromatic image.

We solve the Kohn-Sham equations for this EDF by introducing various external fields to obtain a set of ground state energies 
and nucleon number densities.
The explcit forms of the external fields are specified in Sec. II C. 
We solve the Khon-Sham equations by discretizing the real space, with 
the mesh size of 0.8 fm in both the $r$ and $z$ directions.
We take 10 grid points in the $r$ direction and $20$ points in the $z$-direction, with which the density $\rho(r, z)$ can be considered as a $10\times 20$-dimensional vector in our calculations.
We choose the box boundary condition and include the $z$-component of angular momentum up to $9/2$.

\subsection{Neural network}
In this paper, we carry out a regression analysis of $E = E[\rho]$ using a set of the particle number density and the energy $D =\{E^{(i)}, \rho^{(i)}\}_i$ generated by the KS-DFT.
To this end, we utilize a neural network with fully-connected layers 
for the fitting function.
The fundamental structure of a neural network involves a repetition of linear and nonlinear transformations on the input vector; fully-connected layers signify that all the neurons in the previous layer are connected to all the neutrons in the next layer. 

\begin{figure}[htbp]
    \begin{center}
    \includegraphics[width=90mm]{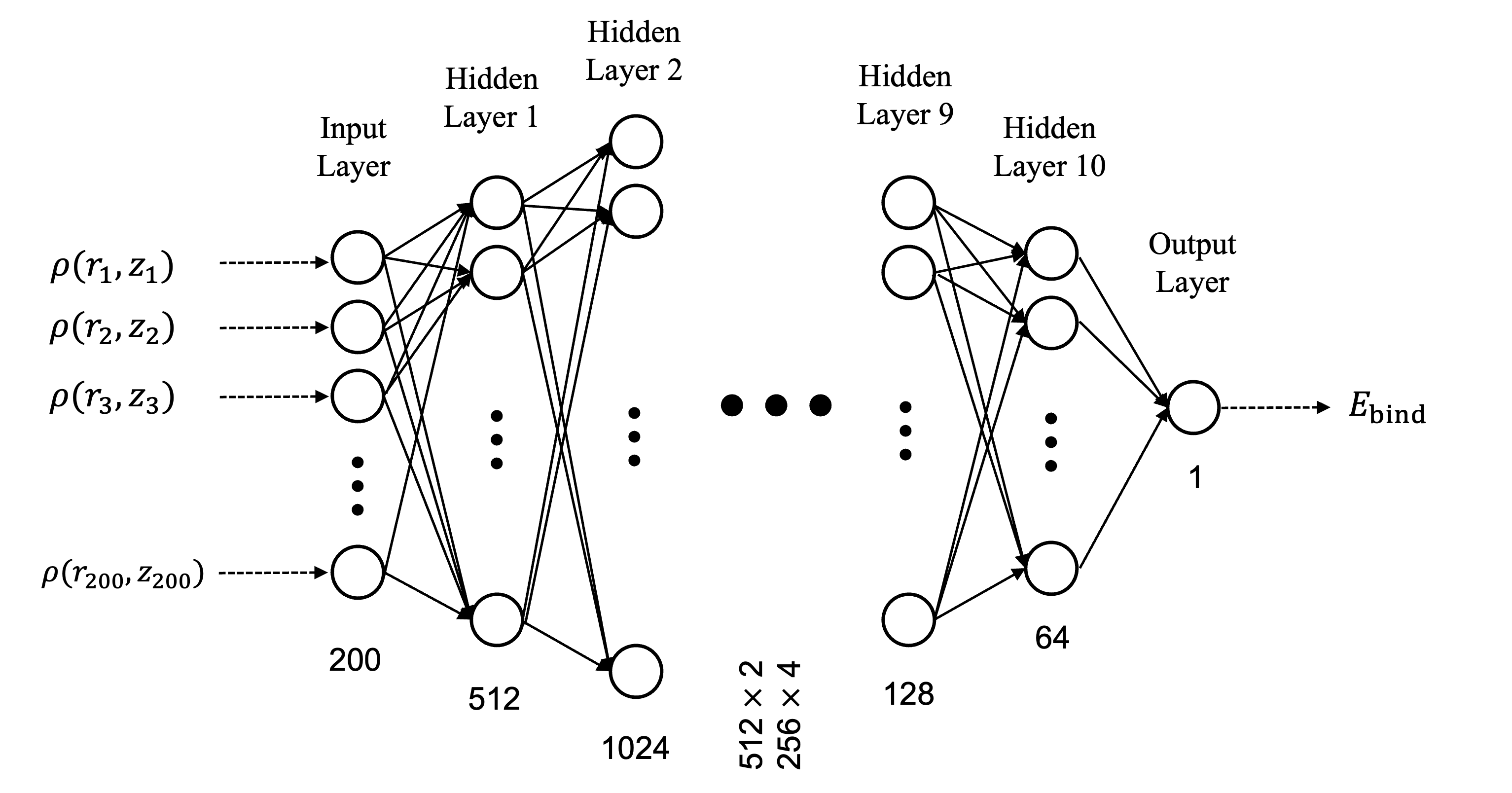}
    \caption{
    \label{fig:DNN}
    A neural network employed in this work to learn the Skyrme-EDF, $E[\rho]$.
    It consists of 10 hidden layers, all of which are fully connected. Their activation functions are the ReLU, and the sigmoid activation function is employed for the output layer.
    The number of neurons in each layer is listed below the layers.
    }
    \end{center}
\end{figure}

We mention that neural networks composed solely of fully-connected layers may 
encounter an issue of an excessive number of parameters when the dimension of an input vector is large. 
To avoid this problem, a convolutional neural network (CNN) is often employed, which has demonstrated a remarkable success in the field 
of computer vision \cite{AlexNet, CNN_review}.
In fact, in the previous application of deep learning to KS-DFT \cite{RS19}, 
the input size of a vector has as large as $256\times 256$ dimension, and thus the CNN was employed. 
However, the dimension of our studies in this paper is much smaller, with $10 \times 20$ dimension. 
Therefore, we do not need to introduce the CNN, and a simpler 
neural network consisting of the fully-connected layers, 
as depicted in Fig. \ref{fig:DNN}, is employed in this study (see the caption for the details).

We use the Adam optimizer \cite{Adam}, which has three tunable parameters. 
Among the three parameters, we set a learning rate to be $10^{-4}$ 
and the others to be default value of the Keras API \cite{keras}.
The batch size is 128, namely we divide training data into subsets, each of which contains 128 components. 
In each update of the fitting parameters, we do it only within each subset to minimize a loss function, for which we take a mean square loss function. 
To avoid the problem of overfitting, we adopt the early stopping strategy and stop the learning at the 500th epoch.
We decrease the learning rate sequentially to $10^{-5}$ (at epoch = 101), $10^{-6}$ (at epoch = 201), $5.0\times10^{-7}$ (at epoch = 301), and $10^{-7}$ (at epoch = 401).

\subsection{External fields}
For a given EDF, one can make a correspondence between the particle number density and the energy of the ground state for a specific external field.
This property will be used to construct a data set to be trained for an OF-EDF. For this purpose, a diverse range of external fields is required.
In this subsection, we introduce two methods to generate the external potentials used in this study. 
The basic idea of these methods is adapted from the previous studies \cite{MS17, RS19} on two-dimensional systems, but we modify them for the axial-symmetric systems.

\subsubsection{Simple Harmonic Oscillators (SHO)}
The first method is to use external fields based on a Simple Harmonic Oscillator (SHO).
As the name implies, this is a deformed harmonic oscillator potential shifted in the $z$-direction:
\begin{equation}
    v^{(i)}_{\mathrm{SHO}}(r,z)
    =
    \frac{1}{2}k^{(i)}_r r^2 
    +
    \frac{1}{2} k^{(i)}_z(z - z^{(i)}_0)^2. 
\end{equation}
The parameters in the range of $0 \leq k_r, k_z \leq 1.1 \, \mathrm{MeV/fm^2}$, and $-1.6\,\mathrm{fm}\leq z_0 \leq 1.6\,\mathrm{fm}$ in the potential are generated from uniform random parameters
\footnote{Unless otherwise noticed, all the random numbers used in this paper are uniform random numbers.}.

The SHO potentials would be able to encompass only a small portion of a domain of the external fields to be used in the Skyrme-EDF.
However, for practical calculations, only a limited variety of external fields, such as a quadrupole moment, has frequently been utilized, if a constrained field is regarded as an external field in a broad sense. 
It is therefore still useful to examine the effectiveness of the learning process with the SHO potentials.

\subsubsection{Random Potentials (RND)}

\begin{figure*}[t]    
    \begin{center}
    \includegraphics[width=175mm]{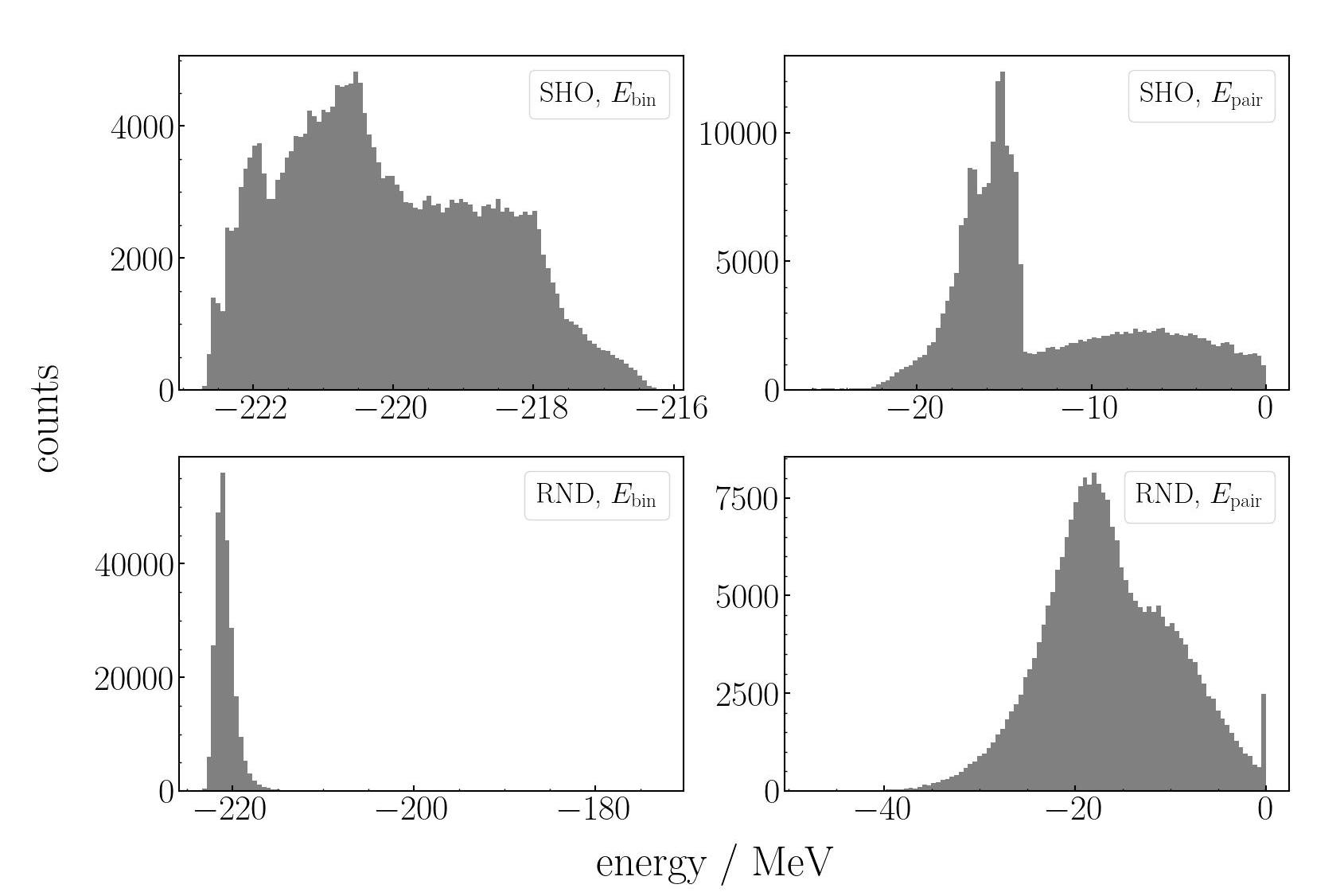}
    \caption{ \label{fig:hist_bin_pair}
    Histograms for the results of the Skyrme EDF calculations with 250,000 different external fields based on the SHO (the top panels) 
    and the RND (the bottom panels) fields.
    The left and right panels show the 
    binding energy and the pairing energy in units MeV, respectively.
    It can be observed that 
    the structure in the shape of the histograms is washed out to a large extent 
    for the RND external fields, which are more random than the SHO cases. 
    }
    \end{center}
\end{figure*}

As the second method, we introduce a Random Potential (RND).
This is a highly random potential with many random numbers:
\begin{equation}
       \label{eq:RND}
    v^{(i)}_{\mathrm{SHO}}(r, z)
    =
    m(r, z)\times\mathrm{sr}^{(i)}(r, z), 
\end{equation}
where $m(r, z)$ and $\mathrm{sr}^{(i)}(r, z)$ are defined as, 
\begin{equation}
 \label{eq:mask}
    m(r, z)
    =
    e^{-4.0\max\{0, \, \sqrt{r^2 + z^2} - r_0\}^2 / r_0^2},
\end{equation}
and 
\begin{equation}
     \label{eq:smoothing}
    \mathrm{sr}^{(i)}(r, z)
    =
    \sum_{r', z'}
    s^{(i)}(r, z ; r', z')\,
    \mathrm{rnd}^{(i)}(r', z'),
\end{equation}
respectively, with  
\begin{equation}
     \label{eq:gaussian}
    s^{(i)}(r, z ; r', z')
    =e^{-\{(r - r')^2 + (z - z')^2\} / \mu_2^{(i)}(r',z')}. 
\end{equation}
The meaning of $\mathrm{rnd}^{(i)}(r, z)$ in Eq. (\ref{eq:smoothing}) and $\mu^{(i)}_2(r,z)$ in Eq. (\ref{eq:gaussian}) is 
as follows. 
First, for each grid point $(r, z)$, a random number within the range of $[v_{\rm min}, v_{\rm max}]$ is generated 
and labeled as $\mathrm{rnd}^{(i)}(r, z)$. 
Since the potential with those random numbers is too irregular to be used as a potential, 
it is smoothed with a Gaussian filter, denoted as $s^{(i)}$, as in Eq. (\ref{eq:smoothing}).
At this stage, the square of the Gaussian width $\mu^{(i)}_2(r,z)$ in Eq. (\ref{eq:gaussian}) is randomly generated within the 
range of $[\mu_{2\rm min}, \mu_{2\rm max}]$ to prevent the external field from acquiring scale information due to the standard deviation of the Gaussian.
Finally, a mask defined by Eq.(\ref{eq:mask}) is applied in Eq.(\ref{eq:RND})
to circumvent a numerical instability caused by a reduction of the external field near the boundary.
In this study, we take 
$r_0 = 1.4\times 1.2A^{1/3}\,\mathrm{fm}$, $v_{\rm min}=-1.1$ MeV, $v_{\rm max}=1.1$ MeV, 
$\mu_{2\rm min}=$ 0.8 fm$^2$, and $\mu_{2\rm max}=$ 1.2 fm$^2$.

In Refs. \cite{MS17, RS19}, random $\{0, 1\}$ binary data were utilized for $\mathrm{rnd}^{(i)}(r, z)$.
For electronic systems, such a choice would be plausible because the potential primarily arises from the Coulomb potential due to a nucleus.
On the other hand, in nuclear systems, it would be a highly non-trivial question to ask which potential is useful to describe static and dynamical properties of atomic nuclei. 
While many calculations employ a phenomenological deformed mean-field potential with e.g., a qudrupole deformation to study deformed nuclei, 
it is not obvious whether such choice is optimal. 
Therefore, in this study, we use random real numbers for $\mathrm{rnd}^{(i)}(r, z)$ to generate more 
diverse external fields than in the previous studies. 
Additionally, since the constraint on the center-of-mass position is included in the definition of the KS-EDF (\ref{eq:KS-EDF}), a different mask function $m$ from that in the previous studies is also introduced. 

\section{Results}

\subsection{Generation of a dataset}

\begin{figure*}[p]
    \begin{center}
    \includegraphics[width=160mm]{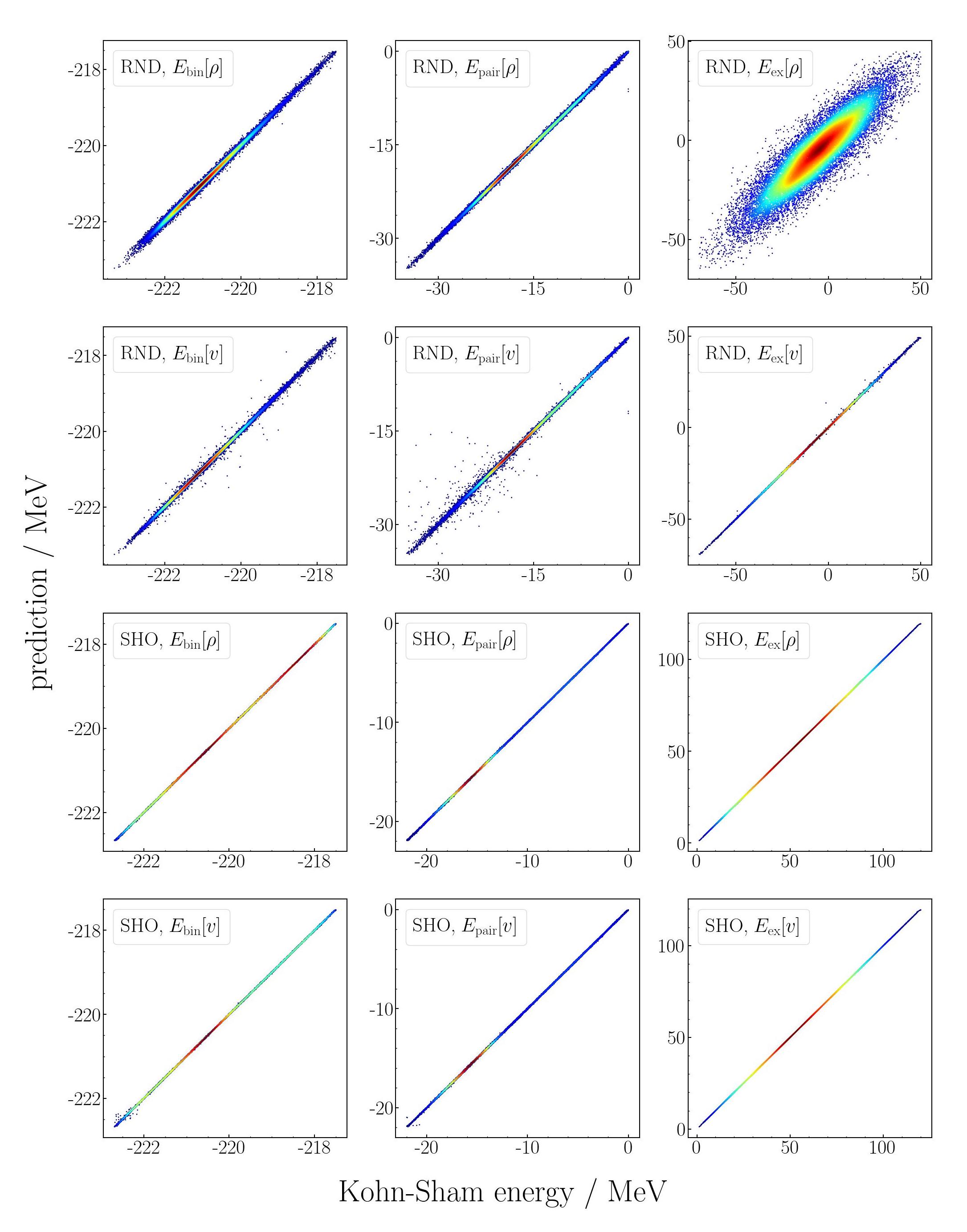}
    \caption{\label{fig:error_bin_pair_ex}
    Comparisons of the Kohn-Sham method (the horizontal axes) and the predicted results from the neural network (the vertical axis) 
    for $E[\rho]$ and $E[v]$ for the RND (the top panels) and SHO (the bottom panels) external fields, all given in units of MeV.
    From the left to the right panels, the training results are shown for the binding energy, the pairing energy, and the energy of the external fields.
    The results for 20,000 test data points are plotted in each figure, in which densely populated (under populated) points are displayed 
    in red (blue). 
    }
    \end{center}
\end{figure*}

Let us now apply the deep learning protocol discussed in the previous section to the Skyrme EDF. 
We first prepare 250,000 data sets for each of the SHO and the RND external fields. 
For each calculation, 
the outputs are 
i) the nucleon number density $\rho$, ii) the kinetic energy $E_{\mathrm{kin}}$, iii) the interaction energy $E_{\mathrm{int}}$, iv) the pairing energy $E_{\mathrm{pair}}$, and 
v) the energy for the external field $E_{\mathrm{ex}}$.
The binding energy $E_{\mathrm{bin}}$ is also computed as a sum of $E_{\mathrm{kin}},~E_{\mathrm{int}}$, and $E_{\mathrm{pair}}$, 
as $E_{\mathrm{bin}}= E_{\mathrm{kin}} + E_{\mathrm{int}} + E_{\mathrm{pair}}$. 
Figure \ref{fig:hist_bin_pair} displays the distribution of $E_{\mathrm{bin}}$ and $E_{\mathrm{pair}}$ for each of the SHO and the RND external fields. 
The distributions of the other components of the energy are summarized in Appendix A (see Fig. \ref{fig:hist_kin_int_ex}).
In order to use these data for deep learning, 
we reject those outside the regions given in Tab. \ref{tab:cutoff}.
From the remaining data, we select 200,000 data for training.
Out of those 200,000 data, we adopt 90 \% of them for training data,  
while the rest for test data, which are not used for training. 

\begin{table}[h]
 \caption{The lower and the upper cut-off energies, in units MeV, for the two different types of the external fields, SHO and RND. 
 For each leraning, only the data within the intervals are employed.
 The value of cutoffs are determined so that approximately all the data shown in Figs. \ref{fig:hist_bin_pair} and \ref{fig:hist_kin_int_ex} can be included.
 }
 \label{tab:cutoff}
 \centering
  \begin{tabular}{c|rr|rr}
   \hline\hline
     & \multicolumn{2}{c|}{SHO} &  \multicolumn{2}{c}{RND} \\
   \hline
   type &   lower   &   upper    & lower & upper \\
   \hline 
   $E_{\rm bin}$  & $-\infty$ & $-$217.5 & $-\infty$ & $-$217.5 \\
   $E_{\rm kin}$ & 395.0 & 450.0 & 360.0 & 420.0 \\
   $E_{\rm int}$ & $-$650.0 & $-$600.0 & $-$630.0 & $-$550.0 \\
   $E_{\rm pair}$ & $-$22.0 & +$\infty$ & $-$35.0 & $+\infty$ \\
   $E_{\rm ex}$ & $-\infty$ & 120.0 & $-$70.0 & 50.0 \\
   \hline\hline
  \end{tabular}
\end{table}

\subsection{$\rho\to E[\rho]$}
We first discuss the results for each energy as an objective variable with the nucleon number density as an explanatory variable.
In other words, we construct the OF-EDF, which yields the energies from a density distribution as an input.
In DFT, apart from an external field, there would be an ambiguity to divide the functional into components:  
$E_{\mathrm{kin}}, E_{\mathrm{int}}$, and $ E_{\mathrm{pair}}$ themselves may not have strict physical meanings.
Nevertheless, these components can be employed as indicators at least for qualitative discussions, and we thus follow Ref. \cite{RS19} to examine the subparts of the EDF.
In particular, it is interesting to investigate the pairing energy $E_{\mathrm{pair}}$, as it qualitatively verifies whether we can learn the effect of superfluidity, or the pair density, with deep learning.
Notice that this was not addressed in the previous study in Ref. \cite{RS19}.

The top panels in Fig. \ref{fig:error_bin_pair_ex} compare the results of the Kohn-Sham method (the horizontal axes) with the neural network predictions (the vertical axes) for the test data with the RND external fields not used in learning. 
The results with the SHO external fields (see the third top panels) are found to be more accurate 
\footnote{This is a natural outcome from the simplicity of the SHO external fields.}.
Figure \ref{fig:error_bin_pair_ex} shows only $E_{\mathrm{bin}}$, $E_{\mathrm{pair}}$, and $E_{\mathrm{ex}}$, 
while the other components of the energy are displayed in Appendix A (see Fig. \ref{fig:error_kin_int}).
If the learning is perfect, the distribution should be diagonal: actually this is almost the case for all the energies except for the energy 
of the external fields plotted in the rightmost figure. 

We have found that the large error in $E_{\rm ex}$  was not improved by changing the learning method, such as a CNN model (see Appendix).
This may be due to the fact that the particle number densities with different external fields tend to have a similar shape because of the saturation property, which results in an information loss in the process of compressing information on the external fields into the density distributions.
Of course, according to the principle of DFT, ideally there should be no loss of information because there is a bijection between an appropriately defined density and an external field.
However, in actual calculations, information on the detailed structure of external fields may be lost due to several numerical errors such as rounding errors, finite difference errors, and errors associated with a convergence criterion in self-consistent calculations.
It is then natural that the prediction error becomes large when one attempts to recover the external field information from such a density distribution.
The inaccuracy in predicting the energy of external fields was reported also in the previous study \cite{RS19}, but the inaccuracy seems more pronounced in atomic nuclei, which are systems with an attractive interaction.
As we will show in the next subsection, this problem can be improved by using the external fields as explanatory variables.

To quantitatively evaluate the errors, we calculate the mean absolute error (MAE) for each learning, which are summarized in Tab. \ref{tab:MAE}. 
It is remarkable that the MAE for the binding energy is as small as 
0.0051 MeV for the SHO external fields and 0.0433 MeV for the RND external fields, which are much more accurate than the accuracy required e.g., 
for a fission barrier of heavy nuclei as well as for nuclear masses. 
For instance, for the latter, the accuracy of 100 keV is required for the r-process studies~\cite{Mumpower:2015ova}.
The MAE for the pairing energy is 0.0233 MeV for the SHO and 0.1567 MeV for the RND. 
These values indicate that the particle number density predicts well the contribution of the pairing correlation, even though the error is slightly larger than that for the binding energy. 

Finally, let us discuss a computational time. 
For the $^{24}$Mg nucleus, it typically takes about a minute to solve the Skyrme-EDF with the Kohn-Sham method and obtain a single training data point.
In marked contrast, the time to predict the energy with the neural network used in this paper from a given density is much shorter, about 0.1 ms. 
The difference in the computational speed will become larger for heavy nuclei. 
This makes a great advantage of using the deep learning method e.g., in plotting a multi-dimensional potential energy surface for nuclear fission studies of heavy nuclei. 

\begin{table}[h]
 \caption{The mean absolute error (MAEs) for each learning with the SHO and the RND external fields. 
 The units are MeV for $E[\rho]$ and $E[v]$, while the MAE for $\rho[v]$ is dimensionless (see Eq. (\ref{eq:MAE})).
 }
 \label{tab:MAE}
 \centering
  \begin{tabular}{c|rr|rr}
   \hline\hline
     & \multicolumn{2}{c|}{SHO} &  \multicolumn{2}{c}{RND} \\
     \hline
   type &   $E[\rho]$   &   $E[v]$   & $E[\rho]$   &   $E[v]$ \\
   \hline \hline
   $E_{\rm bin}$  & 0.0051 & 0.0054 & 0.0433 & 0.0237 \\
   $E_{\rm kin}$ & 0.0165 & 0.0071 & 0.1131 & 0.0900 \\
   $E_{\rm int}$ & 0.0105 & 0.0182 & 0.0431 & 0.1499 \\
   $E_{\rm pair}$ & 0.0233 & 0.0261 & 0.1567 & 0.1411 \\
   $E_{\rm ex}$ & 0.0318 & 0.0105 & 6.6973 & 0.1338 \\
   \hline
    & \multicolumn{2}{c|}{$\rho[v]$} &  \multicolumn{2}{c}{$\rho[v]$} \\
    & \multicolumn{2}{c|}{0.1107} &  \multicolumn{2}{c}{0.4101} \\
    \hline \hline
  \end{tabular}
\end{table}

\subsection{$v \to E[v]$}
While it is somewhat tangential to the topic of DFT, 
there is a certain demand in electronic systems for a functional that directly predicts the energy from a given external field.
Because of this, in the previous study \cite{RS19}, an energy functional $E[v]$ was constructed following the same procedure as that to construct a functional $E[\rho]$. 
Even though it is unclear whether such a functional is useful in nuclear physics, it may be worth investigating whether a functionl $E[v]$ can be 
constructed in connection to the discussion in Ref. \cite{RS19}. 
We therefore carry out similar calculations using the same neural network and dataset as those in the previous subsection, but with the external fields as the explanatory variables.

\begin{figure}[t]
    \begin{center}
    \includegraphics[width=85mm]{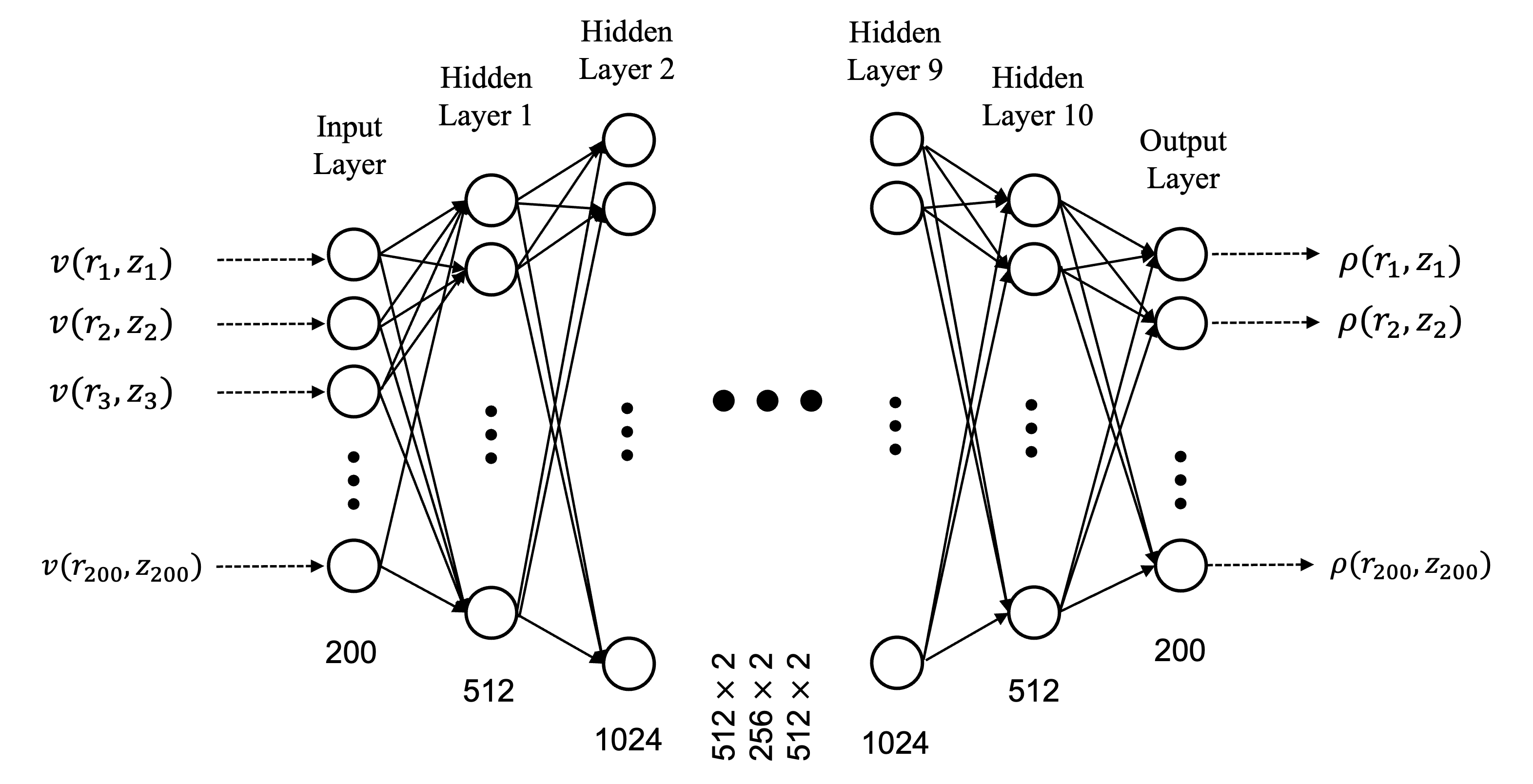}
    \caption{\label{fig:encode}
    A neural network with the encoder-decoder structure employed in this work for a mapping from an external $v$ to a particle number density $\rho$.
    It consists of 10 hidden layers, all of which are fully-connected. Their activation functions are ReLU, 
    and the softmax activation function is employed for the output layer.}
    \end{center}
\end{figure}

The MAEs for $E[v]$ are summarized in Tab. \ref{tab:MAE}, which  
shows that the MAE for $E[v]$ tends to be decreased compared to that for $E[\rho]$.
This is because the external field contains more information than the density distribution.
This is particularly true for learning the energy from the external fields.
On the other hand, the accuracy gets lowered for the binding energy with the SHO external fields. 
To investigate the origin for this, the lower panels in Fig. \ref{fig:error_bin_pair_ex} show comparisons between $E[v]$ from the Skyrme KS calcaulations and the result of the deep learning.  
We find that the points with large errors are due to external fields that have small amplitudes, that is, almost flat potentials. 
Since many SHO potentials used in the dataset have a large curvature, 
it is diffult to learn information about external fields with a small curvature.
Such a problem is less likely to occur in fermionic systems when 
density distributions are used as the explanatory variables, leading to a somewhat better accuracy.

\begin{figure*}[p]
    \begin{center}
    \includegraphics[width=175mm]{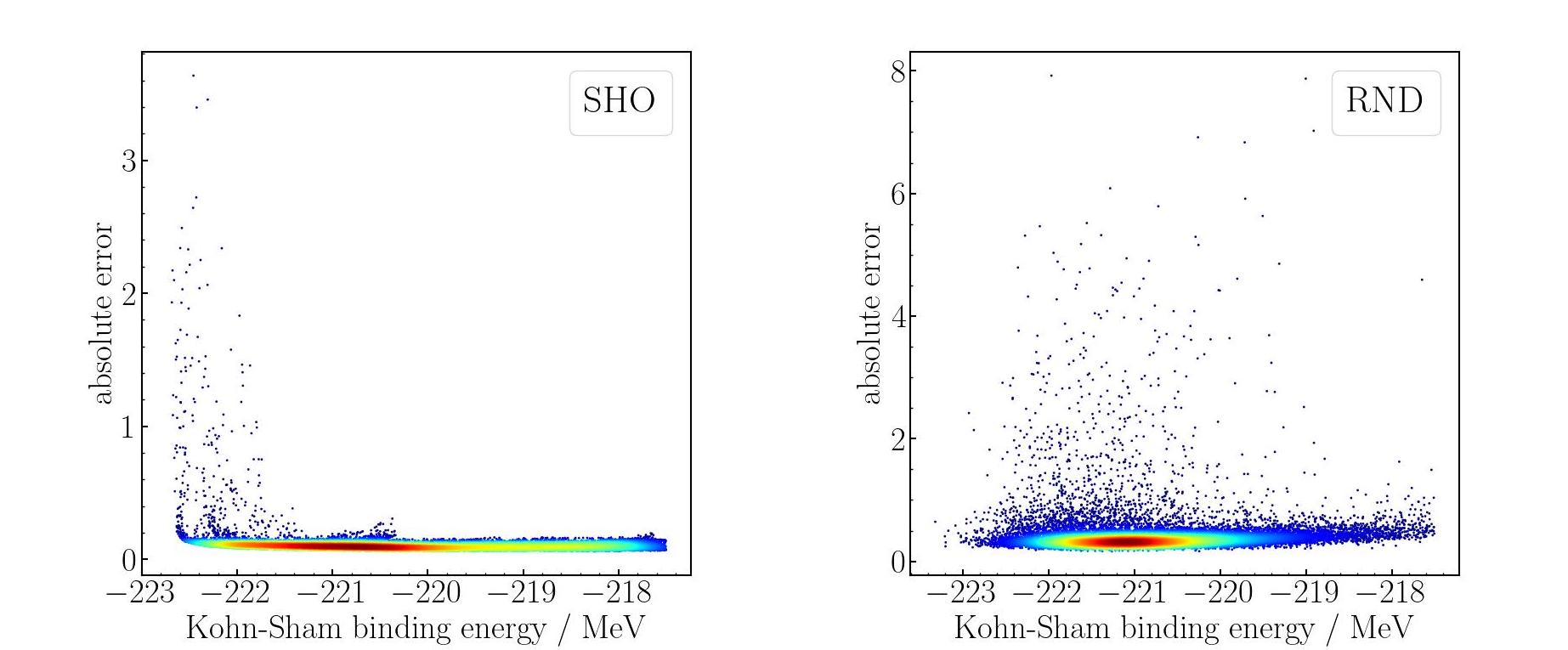}
    \caption{\label{fig:rhov}
    The absolute error of the density distribution directly generated by a deep learning from a given external field $v$. It is 
    plotted as a function of the binding energy from the corresponding Kohn-Sham calculation. 
    The left and the right panels show the results with the SHO and the RND external fields, respectively.
    The densely populated points are displayed in red, while the underpopulated points are shown in blue.
    }
    \end{center}
\end{figure*}

\begin{figure*}[p]
    \begin{center}
    \includegraphics[width=180mm]{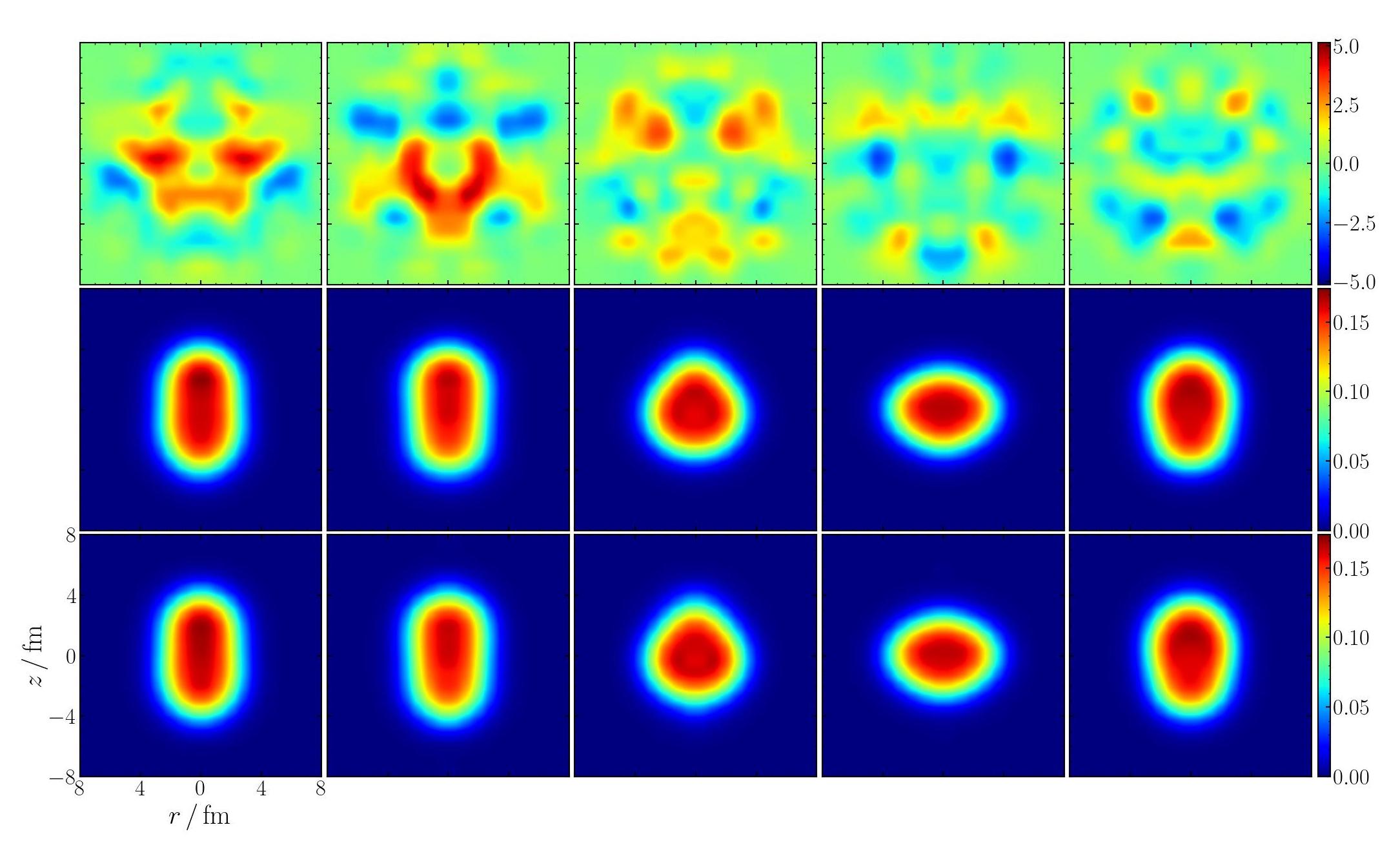}
    \caption{\label{fig:rhov_sample}
    Examples of the predicted densities (the bottom panels) generated directly from the RND external potentials shown in the top panels. 
    For a comparision, the corresponding Khon-Sham densities are also plotted in the middle panels. 
    The units of the color coordinate are MeV for the external potentials and $\mathrm{fm}^{-3}$ for the densities.
    In each panel, the horizontal axis denotes the $r$ coordinate while the vertical axis denotes the $z$ coordinate, whose scales are shown in the left bottom panel. 
    }
    \end{center}
\end{figure*}

\subsection{$v \to \rho[v]$}
Observables are in general calculated in DFT with a particle number density, 
which is obtained with a given functional.
That is, a functional has to be known in advance in obtaining a particle number density. 
As demonstrated in Ref. \cite{RS19}, if a neural network can directly predict the density for a given external field, 
the calculation speed will be significantly improved.
We therefore carry out deep learning for the nuclear system with the external fields as the explanatory variables and 
density distributions as the objective variables.
To this end, we have to take into account the fact 
that the densities are normalized to the particle number, that is, $\int d^3r\,\rho = A$.
The softmax function, which is commonly used in classification problems, enables one to require the normalization condition.
We shall employ this approach in this study for the output layer. 
For the axial symmetric system, the following relationship exists with a discretized spatial mesh:
\begin{equation}
    \frac{2\pi}{A}\iint rdrdz\,\rho(r, z)
    \simeq
    \sum_{i, j}\rho(r_i, z_j)
    \frac{2\pi r_i\Delta r\Delta z}{A}
    =
    1,
\end{equation}
where $\Delta r = \Delta z = 0.8\,\mathrm{fm}$ are the mesh width.

\begin{figure*}[t]
    \begin{center}
    \includegraphics[width=170mm]{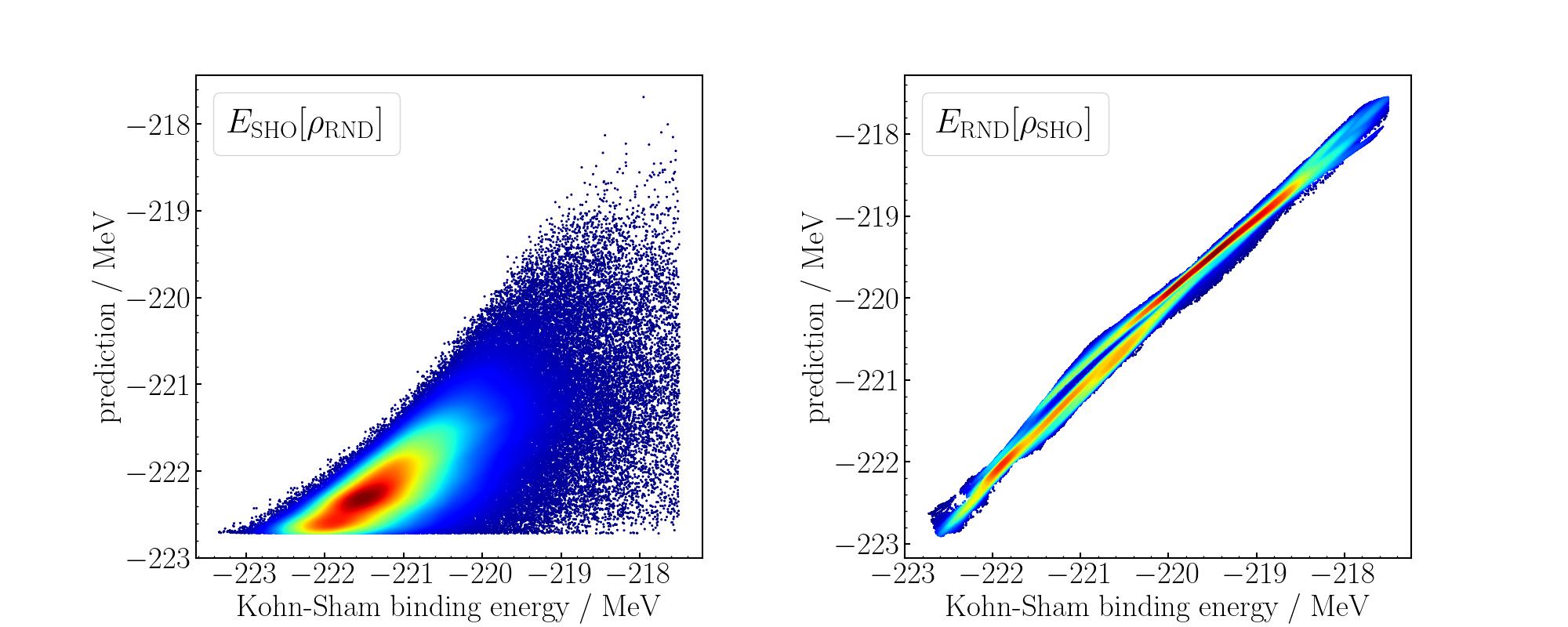}
    \caption{\label{fig:gen}
    A verification of generalization performance for the present deep learning.
    The left and right panels show the results with the RND and the SHO external fields, respectively. 
    The horizontal axes denote the energies obtained with the Kohn-Sham calculations. 
    On the other hand, the vertical axes denote $E_{\rm SHO}[\rho_{\rm RND}]$ (the left panel) and 
    $E_{\rm RND}[\rho_{\rm SHO}]$ (the right panel), 
    that is, 
    the predictions of deep learning trained with the SHO (the left panel) and 
    the RND (the right panel) external fields.  
    Both the taining and test data (200,000 data in total) are plotted in each panels because the RND (SHO) dataset are not used in training $E_{\rm SHO}[\rho]$ ($E_{\rm RND}[\rho])$.
    }
    \end{center}
\end{figure*}

Thus, by selecting $2\pi r_i\Delta r\Delta z\rho(r_i, z_j)/A$ as the objective variable, the normalization is automatically imposed.
In this study, we use a neural network with an encoder-decoder structure for training, as is shown in Fig. \ref{fig:encode}.
The MAE for the learning of $\rho[v]$ is defined as
\begin{equation}
    \label{eq:MAE}
    \mathrm{MAE}
    =
    \overline{2\pi\iint rdrdz\,|\rho_{\mathrm{pred}}(r,z) - \rho_{\mathrm{ans}}(r,z)|}, 
\end{equation}
where $\rho_{\mathrm{pred}}(r,z)$ and $\rho_{\mathrm{ans}}(r,z)$ denote 
a predicted density and a Kohn-Sham result, respectively. 
Here, the bar symbol represents the average over the test data.
We apply the same cut-off energies to the training data as those for the binding energy (see Tab. \ref{tab:cutoff}).

Figure \ref{fig:rhov} shows the error for each test data point plotted as a function of the corresponding binding energy from the Kohn-Sham calculation.
Their average corresponds to the MAE (\ref{eq:MAE}), which is 0.1107 for the SHO external fields and 0.4101 for the RND external fields. 
Figure \ref{fig:rhov_sample} presents the images of the predicted densities 
for a few randomly selected data points for the RND external fields, in comparison to the corresponding Kohn-Sham densities. 
These examples clearly show that our neural networks successfully reproduce the Konh-Sham densities.

\subsection{Generalization performance}
We have so far introduced the two types of external fields and constructed the two independent datasets.
For each dataset, we have successfully provided predictions for the training data with sufficient accuracy; however, 
this does not guarantee performance for unknown data.
For instance, a neural network trained with the RND data does not necessarily yield accurate predictions 
for the SHO data.
This is because the RND and the SHO external fields yield 
density profiles in a different way to each other. 
In general, such generalization performance is a critical concern in applying a trained neural network to another dataset.

To investigate this issue in the context of nuclear physics, 
let us consider $E_{\rm SHO}[\rho_{\rm RND}]$ and $E_{\rm RND}[\rho_{\rm SHO}]$, where $\rho_{\rm SHO}$ and $\rho_{\rm RND}$ are the Kohn-Sham densities obtained with the SHO and the RND external fields, respectively, and 
$E_{\rm SHO}$ and $E_{\rm RND}$ are the functionals trained 
with $\rho_{\rm SHO}$ and $\rho_{\rm RND}$, respectively. 
In Sec. III B, we have investigated $E_{\rm SHO}[\rho_{\rm SHO}]$ and $E_{\rm RND}[\rho_{\rm RND}]$, but 
here we are interested in the performance of the functionals when the densities obtained with the other types of external fields are used as inputs. 
The left panel in Fig. \ref{fig:gen} compares the binding energies 
obtained with the Kohn-Sham calculations with 
the RND external fields with $E_{\rm SHO}[\rho_{\rm RND}]$.
The right panel shows similar quantities, but by inverting RND and SHO, that is 
a comparison between the Kohn-Sham calculations with the SHO external potentials and $E_{\rm RND}[\rho_{\rm SHO}]$.
One can see that the performance of the neural network trained with the SHO external fields, $E_{\rm SHO}$, is quite poor in reproducing the RND test data with large randomness. 
On the other hand, the neural network trained with the RND external fields, $E_{\rm RND}$, successfully predicts the SHO test data, 
although the errors are larger than those for $E_{\rm RND}[\rho_{\rm RND}]$ shown in Fig. \ref{fig:error_bin_pair_ex}. 
The MAEs between Kohn-Sham results and predictions are $1.1523$ MeV for $E_{\rm SHO}[\rho_{\rm RND}]$ and $0.122$ MeV for $E_{\rm RND}[\rho_{\rm SHO}]$.
A similar conclusion has been obtained also in Ref. \cite{RS19}.
Therefore we can conclude that the RND potentials which we adopted are random enough for deep learning.

\section{Summary and future perspectives}
Starting from a Skyrme functional, we have successfully constructed an energy density functional (EDF) 
which depends only on a particle number density.
This functional does not require Kohn-Sham orbitals, and thus can be 
regarded as an orbital-free EDF (OF-EDF).
To this end, we have applied deep learning, in which the density distributions obtained with two types of random external fields (SHO and RND) were mapped on the energy with a neural network. 
The resultant EDF was found to predict various energies for the original Skyrme EDF with reasonable accuracy, except for the energy of the RND external fields, whose accuracy could however be considerably improved when the energies were predicted with deep learning in which the external fields themselves were directly learned. 
The latter feature is more pronounced in systems with an attractive interaction 
than in electron systems.
We have also found that deep learning with less random SHO external potentials has smaller errors as compared to that with the RND external fields.

In this paper, we have employed simple supervised learning.
However, there are various methods of machine learning besides this. 
For example, generative models such as a generative adversarial network (GAN) \cite{GAN, GAN_review} and a diffusion model 
\cite{diffusion_model, stable_diffusion} may provide efficient ways to generate the particle number densities, that is the input for deep learning used in this work to construct an OF-EDF.
These methods maybe useful alternatives for future application of the deep learning method discussed in this paper.

In nuclear physics, a triaxial deformation often plays an important role, particularly in nuclear fission.
In that occasion, one needs to deal with 3-dimensional densities, accounting also for spin and isospin indices.
We mention that traditional neural networks, comprising fully-connected layers, tend not to perform efficiently with such 3-dimensional data, primarily because the data size tends to become huge when the data are converted to 1-dimensional data.
On the other hand, CNNs have shown adaptability to data of general dimensions.
With the Keras API \cite{keras}, 3D CNNs can be conveniently implemented, making a straightfoward extension of the present work to 3D cases. 
Furthermore, the Vision Transformer (ViT) \cite{ViT}, which has recently demonstrated success in image recognition tasks, can also be extended to 3-dimensional data.
With those schemes, the dimensionality of the density itself is not a crucial issue in learning EDFs, without incurring additional costs for preparing training data.

One of the big advantages of using deep learning methods is that energies can be rapidly computed once test data are prepared and trained. 
With such low-cost calculations, numerical experiments will become much easier than before. 
We mention that, as objectives of research become more and more sophisticated, the number of DFT calculations required to publish a single research paper has in general been increased in these days.
A typical example is a calculation for fission barriers in a multi-dimensional space.  
Even though computer performance continues to be improved, 
a computational cost of research has in general been increased, 
and it has been more complicated than before to test an idea with a numerical experiment. 
Fast computational methods like the one developed in this work, particularly when they are provided in a convenient format such as a Python library, 
can significantly shorten the time required to test and validate ideas.
If a numerical accuracy is an issue, one may revalidate ideas obtained with deep learning by using the traditional Kohn-Sham scheme.
This could be interpreted as an application of the idea of materials informatics (MI) \cite{RB17} to a theoretical research.

A potential problem in performing supervised learning is that one has to collect a large set of training data.
In this work, we have chosen a relatively light nucleus, $^{24}$Mg, and imposed axial symmetry, and thus we have treated a relatively low-cost system. 
However, heavy and superheavy nuclei, such as uranium isotopes, will be very costly in terms of data collection, especially when no symmetry is imposed, even though those nuclei have attracted lots of attention in 
nuclear physics, as e.g. a finding the optimal pathway in fission has still been a big theoretical challenge. 
In this regard, we would like to point out that a data collection needs not be performed individually; it could actually be done collaboratively by many researchers.
A lot of good quality data, which are ready to be used in deep learning, may have already existed for some selected nuclei. 
Therefore, we believe that it is desirable to establish a framework in the nuclear theory community to collect numerical data and/or to carry out numerical calculations with unified hyperparameters such as a mesh size. 
Such a collaborative approach will help advance research more efficiently and effectively, benefiting the whole nuclear physics community.

\section*{Acknowledgments}
We thank G. Col\`o for useful discussions. 
This work was supported by JSPS KAKENHI
(Grant Nos. 21J22348, JP19K03824, JP19K03861, JP19K03872, and JP23K03414).

\bibliographystyle{apsrev4-2}
\bibliography{ref}

\appendix


\section{Figures for $E_{\rm kin}$ and $E_{\rm int}$}

In this Appendix, we show figures for $E_{\rm kin}$ and $E_{\rm int}$ which are not shown in Sec. III. We also show a 
figure for a histogram for $E_{\rm ex}$. The conclusions remain the same as those for Figs. 2 and 3.

\begin{figure*}[p]    
    \begin{center}
    \includegraphics[width=175mm]{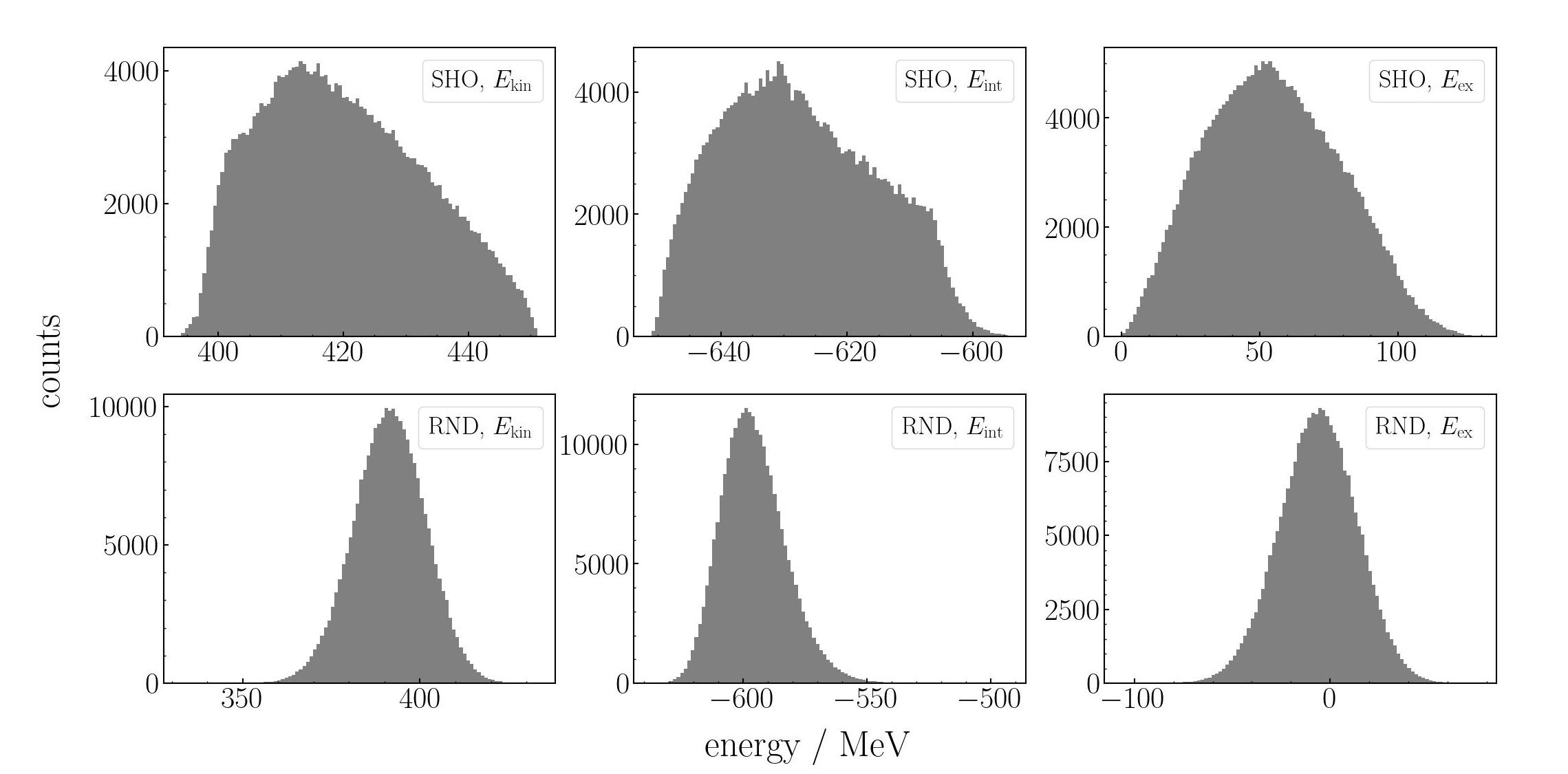}
    \caption{\label{fig:hist_kin_int_ex}
    Same as Fig. 2, but for the 
     kinetic energy, the interaction energy, and the energy for the external field.
    }
    \end{center}
\end{figure*}

\begin{figure*}[p]
    \begin{center}
    \includegraphics[width=120mm]{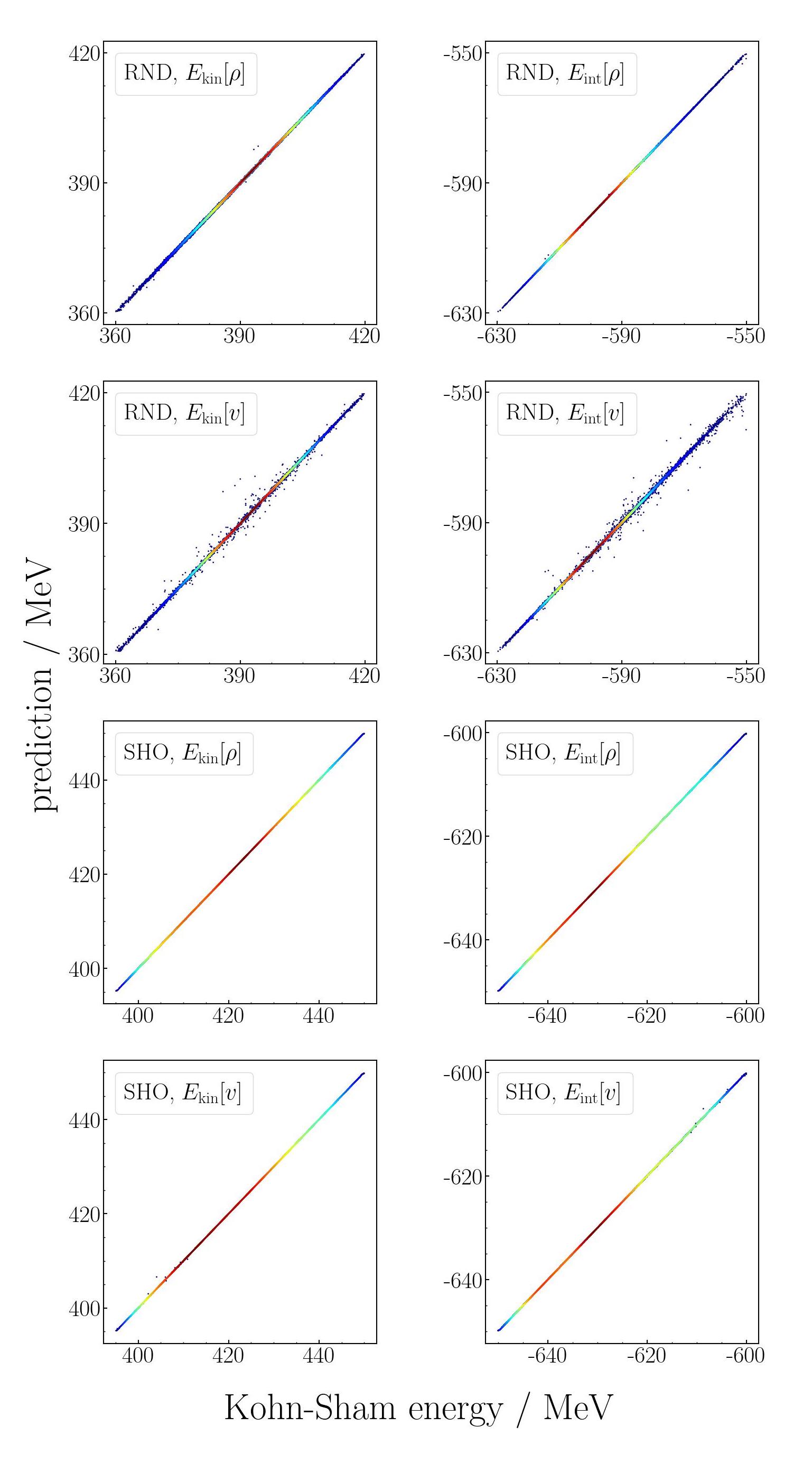}
    \caption{\label{fig:error_kin_int}
    Same as Fig. 8, but for the kinetic and the interaction energies.
    }
    \end{center}
\end{figure*}

\clearpage

\section{Convolutional Neural Network}

In the previous study \cite{RS19}, the convolutional neural network (CNN) was used for calculating an OF-EDF.
An efficient structure in the CNN enables computers to recognize images, which reduces the trainable parameters, and the CNN works well in learning with large-size images.
Since the inputted image size is small enough in this study, we can use the neural network only consisting of the fully-connected layer.
We can in principle perform the training with CNNs as well, but we find that the results are not significantly 
different (see Table V).
To this end, we use the CNN listed in Tabs. \ref{tab:CNN} and \ref{tab:CNN_encode}.
The learning methods are the same as for the fully-connected layer. 
Considering that the CNN is computationally more expensive than the fully-connected layer, the benefit of using the CNN does not seem to be substantial, at least for the system studied in this paper.

Of course, there are many choices and hyperparameters in deep learning. 
Therefore, we do not mean to claim that there is no sufficient benefit from using CNNs.
However, if one needs better and more accurate training results, skill and experience are required.
In that occasion, one way to proceed is to ask professionals to submit their ideas in a competition on sites such as Kaggle \cite{kaggle}, for example. 
Hopefully, such approach could help identify optimal machine learning methods and hyperparameters.
For example, the IceCube held a competition on Kaggle \cite{IceCube}.

\begin{table*}[t]
\caption{A CNN employed in this work to learn $E[\rho]$ and $E[v]$.
The type of each layer is expressed with the language of Keras API \cite{keras}, with which one can reproduce this neural network easily.
In each layer, all arguments not mentioned are default values.
 }
 \label{tab:CNN}
 \centering
  \begin{tabular}{cc}
  \hline\hline
  layer & type \\
  \hline
  input & Input(shape=(10, 20, 1)) \\
  1 & Conv2D(filters=32, kernel\_size=3, activation='relu') \\
  2 & Conv2D(filters=64, kernel\_size=3, activation='relu') \\
    
  3--8 & Conv2D(filters=64, kernel\_size=4, padding='same', activation='relu') \\

  9 &  Conv2D(filters=128, kernel\_size=3, activation='relu') \\

  10--13 & Conv2D(filters=128, kernel\_size=3, padding='same', activation='relu') \\
  14 & Flatten() \\
  15 & Dense(units=128, activation='relu') \\
  output & Dense(units=1, activation='sigmoid') \\
  \hline\hline
  \end{tabular}
\end{table*}

\begin{table*}[p]
\caption{A CNN employed in this work to learn $\rho[v]$.
The neural network has encoder-decoder structure.
Layer 1--4 are reducing convolutional layers, and Layer 5--7 are non-reducing convolutional layers, where the size of images are (2, 12).
We use Layer 8--11 as deconvolution layers to enlarge them up to (10, 20).
In each layer, all arguments not mentioned are default values.
}

 \label{tab:CNN_encode}
 \centering
  \begin{tabular}{cc}
  \hline\hline
  layer & type \\
  \hline
  input & Input(shape=(10, 20, 1)) \\
  1 & Conv2D(filters=32, kernel\_size=3, activation='relu') \\
  2 & Conv2D(filters=64, kernel\_size=3, activation='relu') \\
  3 & Conv2D(filters=128, kernel\_size=3, activation='relu') \\
  4 & Conv2D(filters=256, kernel\_size=3, activation='relu') \\
    
  5--7 & Conv2D(filters=256, kernel\_size=4, padding='same', activation='relu') \\

  8 &  Conv2DTranspose(filters=256, kernel\_size=3, activation='relu') \\
  9 &  Conv2DTranspose(filters=128, kernel\_size=3, activation='relu') \\
  10 &  Conv2DTranspose(filters=64, kernel\_size=3, activation='relu') \\
  11 &  Conv2DTranspose(filters=32, kernel\_size=3, activation='relu') \\
  12 & Flatten() \\
  13 & Dense(units=200, activation='softmax') \\
  output & Reshape(target\_shape=(10, 20, 1)) \\
  \hline\hline
  \end{tabular}
\end{table*}

\begin{table*}[h]
 \caption{The mean absolute errors (MAEs) for each learning with the SHO and the RND external fields using the CNNs. 
 The units are MeV for $E[\rho]$ and $E[v]$, while the MAE for $\rho[v]$ is dimensionless.
 }
 \label{tab:CNN_MAE}
 \centering
  \begin{tabular}{c|rr|rr}
   \hline\hline
     & \multicolumn{2}{c|}{SHO} &  \multicolumn{2}{c}{RND} \\
     \hline
   type &   $E[\rho]$   &   $E[v]$   & $E[\rho]$   &   $E[v]$ \\
   \hline \hline
   $E_{\rm bin}$  & 0.0049 & 0.0055 & 0.0336 & 0.0245 \\
   $E_{\rm kin}$ & 0.0151 & 0.0079 & 0.1010 & 0.1134 \\
   $E_{\rm int}$ & 0.0119 & 0.0191 & 0.0484 & 0.1619 \\
   $E_{\rm pair}$ & 0.0179 & 0.0250 & 0.1505 & 0.1852 \\
   $E_{\rm ex}$ & 0.0220 & 0.0108 & 4.7059 & 0.0889 \\
   \hline
    & \multicolumn{2}{c|}{$\rho[v]$} &  \multicolumn{2}{c}{$\rho[v]$} \\
    & \multicolumn{2}{c|}{0.1092} &  \multicolumn{2}{c}{0.3484} \\
    \hline \hline
  \end{tabular}
\end{table*}

\end{document}